\documentclass[10pt]{iopart}
\usepackage{iopams}
\usepackage{theorem}
\usepackage[dvips]{graphicx,epsfig}
\usepackage{amssymb,bbm}
\usepackage{euscript}
\usepackage{color}

\begin{document}

\title{Distributed coherent manipulation of qutrits by virtual excitation processes}
\author{Zhen-Biao Yang,$^1$ Sai-Yun Ye,$^{1,2}$ Alessio Serafini,$^2$
Shi-Biao Zheng$^{1,*}$}
\address{$^1$Department of Physics and State Key Laboratory Breeding Base of
Photocatalysis, Fuzhou University, Fuzhou 350002, P. R. China\\
$^2$Department of Physics \& Astronomy, University College London, Gower
Street, London WC1E 6BT, United Kingdom}
\address{$^*$Corresponding author: sbzheng@pub5.fz.fj.cn}

\begin{abstract}
We propose a scheme for the deterministic coherent manipulation of
two atomic qutrits, trapped in separate cavities coupled through a
short optical fibre or optical resonator. We study such a system in
the regime of dispersive atom-field interactions, where the dynamics
of atoms, cavities and fibre operates through virtual population of
both the atomic excited states and photonic states in the cavities
and fibre. We show that the resulting effective dynamics allows for
the creation of robust qutrit entanglement, and thoroughly
investigate the influence of imperfections and dissipation, due to atomic spontaneous
emission and photon leakage, on the entanglement of the two qutrits
state.
\end{abstract}


\maketitle

\section{Introduction.}

One of the crucial ingredients in the upcoming area of quantum
technologies will be the capability of coherently manipulating
quantum systems at a distance, such that {\em entanglement} ({\em
i.e.}, quantum correlations) can be created between different nodes
of a global quantum system.

Entanglement is one of the most peculiar features of quantum
mechanics, and the most distinct signature of quantum coherence.
Entangled states of two or more particles not only play an important
role for tests of quantum nonlocality [1-3], but also lie at the
heart of quantum information processing and quantum computing [4]. Entangled
quantum states come in many flavours, such as Bell,
Einstein-Podolsky-Rosen [1], Greenberger-Horne-Zeilinger [3], or W
states [5], generally depending on the dimensionality and tensor
product structure of the Hilbert spaces involved. All these states
have different qualities and are suitable for different roles in
quantum information protocols [3,5]. In this context, entangled
states of multiple systems with Hilbert spaces of dimension $d$
({\em i.e.} of `qudits', with $d>2$) offer their specific
advantages over the -- archetypical and most commonly considered --
entangled states of two-dimensional systems (of `qubits'). For
instance, entangled states of two qudits
violate local realism more strongly than entangled states of two
qubits, and their entanglement is more resilient to noise [6].
Also, quantum cryptographic protocols where qubits are replaced with
qudits are both more secure and faster (in that more information may
be sent, on average, per sent particle) [7].

Entangled states can currently be generated in a variety of physical
systems, such as trapped ions [8], quantum electrodynamics cavities
(QED) [9], superconducting circuits [10], semiconductor quantum dots
[11], linear optical systems [12] and impurity spins in solids [14].
Cavity QED [14], which concerns the interaction of atoms and photons
inside optical cavities, provides experimentalists with a very
favorable setting for the generation of entanglement. Atomic systems
are qualified to act as qubits or, more generally, qudits, as
appropriate internal electronic states can coherently store
information over relatively long time scales. At the same time, in
such systems photons are suitable for the transfer of information
between distant nodes. High-finesse cavities can provide good
insulation against the environment and can thus have long coherence
times [15]. Two-atom Bell states, and three-particle (two atoms plus
one photon) GHZ entangled states have been experimentally
demonstrated with Rydberg atoms passing through a superconducting
microwave cavity [16,17].

Schemes for the generation of entangled states of two {\em
`qutrits'} ({\em i.e.,} of two three level quantum systems) for two
atoms via a single nonresonant cavity have also been proposed [18].
However, in order to be used for quantum communication protocols
[19], such an entanglement should be generated between {\em
distant} atoms, like atoms trapped in different cavities.
``Distributed'' atomic entanglement requires a way to coherently mediate
the interaction between the two atoms. One way to establish this
interaction is through coincidence detection of photons leaking out
of the cavities [20]: in this way, entangled states are only
probabilistically generated and the success probability is dependent
on the efficiency of photon detectors. The other possibility is to
directly link the cavities with an optical fibre, waveguide or by a
third mediating cavity: entangled states can be
deterministically generated in such a way [21,22]. Both types of quantum
connectivity essentially allow for the distribution of entanglement
across a quantum network [23].

In this paper, we shall present a way, based on the proper choice of
atomic levels' structure and operating regimes, to engineer a
deterministic coherent interaction between two qutrits embodied by
atoms trapped in distant cavities, linked by a third optical
resonator. We will then move on to study the entanglement that can
be generated by such an interaction, as well as its resilience to
imperfections and quantum noise.

Before proceeding, let us first review some previous schemes for the
deterministic generation of entangled states via such a type of
connected cavities [22]. Several schemes have been proposed for the
deterministic generation of several diverse kinds of entangled states
[24-28], including Bell states [24], W states [28], GHZ states
[26,28], and also qutrit entangled states [25,27]. In the schemes of
Refs. [25,26], the adiabatic passage along dark states is employed.
These schemes [25,26] are based on accurately tailored sequences of
pulses and thus require a considerable degree of control. In other
schemes [22,24], the Rabi oscillations of the whole system composed
of the atoms, cavity modes, and fibre modes is utilized; the
entangled states are generated through the exchange of excitation
numbers for the atoms and photons. Hence, such schemes [22,24] are
bound to be rather sensitive to atomic spontaneous emission and
photon losses. In other schemes [27,28], the population of the atomic
excited states can be effectively suppressed by virtue of dispersive
atom-field interactions, but the photonic states in the cavities or
fibre are still populated, so that the whole system is still
sensitive to photon losses.

The scheme we propose here is different from all such previous
schemes [22,24-28], even from those adopting adiabatic passage
[25,26]. The scheme is inspired by a previous idea for
virtual-photon-induced phase gates between two distant atoms [29].
In the scheme, the entanglement is generated through virtual
population of not only the atomic excited states but also the
photonic states in the cavities and fibre. Therefore, our scheme
will turn out to be well shielded from both atomic spontaneous
emission and photon losses.


The paper is organized as follows. In Sec.~\ref{model}, we specify
our conditions on the physical parameters and derive the effective
Hamiltonian for the system.
In Sec.~\ref{entanglement}, we discuss the generation of two qutrit
entangled state via the effective Hamiltonian and study the
reliability of the entangled state in the presence of mismatches in
the system's parameters. In Sec.~\ref{noise}, we discuss the
influence of atomic spontaneous emission and photon leakage on the
entangled state. Sec.~\ref{conclu} contains some concluding remarks.

\section{The model}\label{model}

\begin{figure}[t!]
\centerline{
\includegraphics[scale=.5]{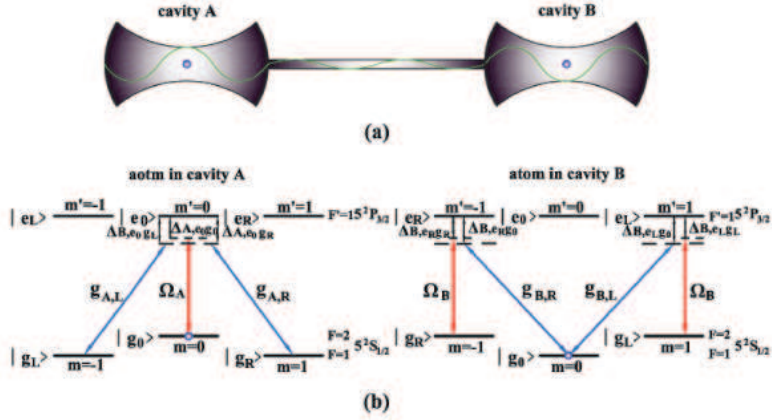}}
\caption{Setup and atoms' levels' configuration for realising qutrit
entanglement. (a) Two atoms are trapped in the double-mode cavities
A and B, respectively; the cavities are coupled by an optical fibre.
(b) Possible implementation with $^{87}$Rb atoms, showing the
involved atomic transitions for each atom.}
\end{figure}

The schematic of our setup is shown in Fig.~$1$. Two distant atoms
are individually trapped in two double-mode cavities (A and B),
which are connected by a third optical resonator, as shown in
Fig.~$1$ (a). The linking resonator can be either a third cavity
coupling the two distant cavities (like in a photonic crystal), or a
`short' (in a sense which will be specified shortly) optical fibre.
For simplicity, we will henceforth refer to the linking resonator as
to the ``fibre''.

The coupling of the fibre modes to the modes of the cavities in Schr\"odinger
interaction picture may be modeled by the Hamiltonian $H_I^{cf}=\sum_{n=1}^%
\infty \sum_{k=L,R}\Delta _{n,k}b_{n,k}^{\dag}b_{n,k}+\nu
_{n,k}\{b_{n,k}[a_{A,k}^{\dag}+(-1)^ne^{i\varphi _f}a_{B,k}^{\dag})]+H.c.\}$ ($%
\hbar =1$ is used throughout this paper), where $\Delta _{n,k}$ is
the frequency difference of the $n$th polarised fibre mode and the
cavity mode with the corresponding polarisation ($L$ and $R$ denote,
respectively, $\sigma ^{\dag}$-circular and $\sigma ^{-}$-circular
polarisation), $b_{n,k}$ and $a_{A,k}$ $(a_{B,k})$ are the
annihilation operators for the polarised modes of the fibre and of
cavity A (B), $\nu _{n,k}$ is the corresponding coupling strength,
and the phase $\varphi _f$ is due to the propagation of the field
through the fibre of length $L$: $\varphi =2\pi \omega L/c$ [22].
In the short fibre limit $2L\overline{\nu }/(2\pi c)\ll 1$ [22], where $%
\overline{\nu }$ is the decay rate of the cavity fields into a
continuum of the fibre modes, only the resonant modes $b_L$ and
$b_R$ of the fibre are excited and coupled to the cavity modes. In
this case, the interaction Hamiltonian $H_I^{cf}$ describing the
cavity-fibre coupling can be rewritten as [22,27]
\begin{equation}
H_I^{cf}=\sum_{k=L,R}\nu _k[b_k(a_{A,k}^{\dag}+e^{i\varphi
_f}a_{B,k}^{\dag})+H.c.].
\end{equation}
In this paper, the state of the photon modes for cavity A (B) or the fibre
is taken to be $\left| ii^{\prime }\right\rangle _s$ ($s=c_1$, $c_2$, and $%
fib$), with $i$ ($i^{\prime }$) denoting $i$ $\sigma ^{\dag}$- ($i^{\prime }$ $%
\sigma ^{-}$-) photons.

The atoms have three excited states ($\left| e_L\right\rangle $, $\left|
e_0\right\rangle $, and $\left| e_R\right\rangle $) and three ground states (%
$\left| g_L\right\rangle $, $\left| g_0\right\rangle $, and $\left|
g_R\right\rangle $), which could be the Zeeman sublevels of alkali atoms in
the excited- and ground-state manifold, respectively.
To fix ideas and portrait a case of practical interest, we consider here a
possible implementation with $^{87}Rb$, whose relevant atomic levels are shown
in Fig.~$1.$ (b). We only illustrate the involved state transitions by
starting from the initial state $\left| g_0\right\rangle _A\left|
g_0\right\rangle _B$ for the atoms. Each atom is assumed to be coupled
(off-resonantly) to an external $\pi $-polarised classical field and both $%
\sigma ^{\dag}$- and $\sigma ^{-}$-polarised photon modes of the local cavity.

We first describe the involved transitions for each atom in its local
cavity. In cavity A, the transitions $\left| g_0\right\rangle \rightarrow
\left| e_0\right\rangle $ and $\left| e_0\right\rangle \rightarrow \left|
g_L\right\rangle $ ($\left| g_R\right\rangle $) are coupled to the $\pi $%
-polarised classical field and the $\sigma ^{\dag}$-circular ($\sigma ^{-}$%
-circular) polarised cavity mode, respectively. In cavity B, the transitions
$\left| g_0\right\rangle \rightarrow \left| e_L\right\rangle $ ($\left|
e_R\right\rangle $) and $\left| e_k\right\rangle \rightarrow \left|
g_k\right\rangle $ are coupled to the $\sigma ^{\dag}$-circular ($\sigma ^{-}$%
-circular) polarised cavity mode and the $\pi $-polarised classical field,
respectively. It should be noted that, for the selected frequencies of the
classical and cavity fields, additional transitions cannot be induced due
to the large difference of the energy levels between the $F=1$ and $F=2$ states
of the ground manifold $5^2S_{1/2}$. In interaction picture, the
Hamiltonian describing the interaction of the atoms with the cavity and
classical fields can then be written as
\begin{eqnarray}
\fl H_I^{acl} &=&\sum_{k=L,R}[g_{A,k}a_{A,k}e^{i\Delta _{A,e_0g_k}t}\left|
e_0\right\rangle _A\left\langle g_k\right| +\Omega _Ae^{i\Delta
_{A,e_0g_0}t}e^{i\phi _A}\left| e_0\right\rangle _A\left\langle g_0\right|
\nonumber \\
\fl && + g_{B,k}a_{B,k}e^{i\Delta _{B,e_kg_0}t}\left| e_k\right\rangle
_B\left\langle g_0\right| +\Omega _Be^{i\Delta _{B,e_kg_k}t}e^{i\phi
_B}\left| e_k\right\rangle _B\left\langle g_k\right| +H.c],
\end{eqnarray}
where $\Delta _{x,yz}$ ($x=A,B;$ $y=e_0,e_L,e_R;$ $z=g_0,g_L,g_R$) denotes
the energy difference between the fields and the corresponding atomic
transition $\left| y\right\rangle \leftrightarrow \left| z\right\rangle $ in
cavity $x$; $g_{x,k}$ is the coupling strength of the atom with the
polarised photon mode in cavity $x$ and satisfies $g_{x,k}=g_0C_{m,m^{\prime
}}$ (with $g_0$ and $C_{m,m^{\prime }}$ being the atom-cavity coupling
constant and Clebsch-Cordan coefficient, respectively); $\Omega _x$ and $%
\phi _x$ are one-half Rabi frequency and phase of the classical field, and $%
H.c$ denotes Hermitian conjugate.

Under the condition of larger detuning, i.e., $\Delta _{x,yz}\gg $
$g_{x,k}$, $\Omega _x$, the probability that the excited atomic
states are populated is virtual, then the Hamiltonian $(2)$ is
reduced to an effective one that involves only the Stark shifts
induced respectively by the classical and cavity fields for the
three ground states, and Raman transitions $\left| g_0\right\rangle
\rightarrow \left| g_k\right\rangle $ $(k=L,R)$ induced collectively
by the classical and cavity fields (See Appendix A). Furthermore, to
avoid excitation of real photonic states in the cavities and fiber,
we first set $\mu _1\equiv \frac{\Omega _A^2}{\Delta
_{A,e_0g_0}}=\frac{\Omega _B^2}{\Delta _{B,e_kg_k}}$, $\mu _2\equiv
\frac{g_{A,k}^2}{\Delta _{A,e_0g_k}}=\frac{g_{B,k}^2}{\Delta
_{B,e_kg_0}}$, $\lambda \equiv \frac{g_{A,k}\Omega _A}2(\frac
1{\Delta _{A,e_0g_0}}+\frac 1{\Delta
_{A,e_0g_k}})=\frac{g_{B,k}\Omega _B}2(\frac 1{\Delta
_{B,e_kg_0}}+\frac 1{\Delta _{B,e_kg_k}})$, $\Delta \equiv \Delta
_{A,e_0g_k}-\Delta _{A,e_0g_0}=\Delta _{B,e_kg_0}-\Delta
_{B,e_kg_k}$,  $\phi _A=\phi _B$ and $\nu =\nu _k$, and satisfy the
condition $\sqrt{2}\nu $, $\left| \Delta -\sqrt{2}\nu \right| $,
$\Delta +\sqrt{2}\nu $, and $\Delta \gg \frac{\mu _2}4$, $\frac
\lambda 2$. In this case, the energy exchange between the atoms and
the photonic modes of the cavities and fiber is also virtual (see
Appendix A). Suppose all the modes of the cavities and fiber are
initially in the vacuum state, i.e., $\left| 00\right\rangle
_{c_1}\left| 00\right\rangle _{fib}\left| 00\right\rangle _{c_2}$.
Thus all these modes will remain in the vacuum state during the
evolution. Therefore, the global effective Hamiltonian reads [29]
\begin{equation}
\fl H_e^{\prime \prime }=\sum_{k=L,R}\eta (\left| g_0\right\rangle
_A\left\langle g_0\right| +\left| g_k\right\rangle _B\left\langle
g_k\right| )-\chi (e^{-i\varphi _f}S_{A,k}^{\dag}S_{B,k}^{-}+H.c),
\end{equation}
where
\begin{equation}
\eta =\mu _1+\frac{\lambda ^2}4[(\frac 1{\Delta -\sqrt{2}\nu }+\frac 1{\Delta +%
\sqrt{2}\nu }+\frac 2\Delta ),
\end{equation}
\begin{equation}
\chi =\frac{\lambda ^2}4(-\frac 1{\Delta -\sqrt{2}\nu }-\frac 1{\Delta +%
\sqrt{2}\nu }+\frac 2\Delta ),
\end{equation}
$S_{A,k}^{\dag}=\left| g_0\right\rangle _A\left\langle g_k\right| $, and $%
S_{B,k}^{-}=\left| g_0\right\rangle _B\left\langle g_k\right| $. The
Hamiltonian $(3)$ allows for the global coherent manipulation of the
atomic states. We will show this in detail by studying the
generation of qutrit entanglement between the two distant atoms.

\section{{Generation of qutrit entanglement}}\label{entanglement}

We now show that the effective Hamiltonian $(3)$ allows one to
generate a qutrit-qutrit entangled state between two atoms $A$ and
$B$. Initially, the two cavities and the fibre are in the vacuum
state while the two atoms are initialised in $\left| \psi
_{AB}(0)\right\rangle \equiv \left| g_0\right\rangle _A\left|
g_0\right\rangle _B$ (this can be achieved by optical pumping with
two classical laser fields, one at resonance with the transition
from $F=2$ to $F^{\prime }=2$, the other one coupling to the
transition from $F=1$ to $F^{\prime }=2$ [30]). From the effective
Hamiltonian $(3)$, we immediately obtain the temporal evolution of
the two atoms as follows:
\begin{eqnarray}
\fl \left| \psi _{AB}(t)\right\rangle &=&e^{-i\mu t}[\cos (\sqrt{2}\chi t)\left|
g_0\right\rangle _A\left| g_0\right\rangle _B  \nonumber \\
\fl&& +\frac i{\sqrt{2}}e^{-i\varphi _f}\sin (\sqrt{2}\chi t)(\left|
g_L\right\rangle _A\left| g_L\right\rangle _B+\left| g_R\right\rangle
_A\left| g_R\right\rangle _B)].
\end{eqnarray}
Setting $\chi t=\frac{\arctan (\sqrt{2})+m\pi }{\sqrt{2}}$ $(m=0,1,2,...)$,
we get a qutrit-qutrit maximally entangled state (in the sense that the local
Von Neumann entropy is maximal)
\begin{equation}
\fl \left| \psi _{AB}^{3D}\right\rangle =\frac{e^{-i\mu t}}{\sqrt{3}}[\left|
g_0\right\rangle _A\left| g_0\right\rangle _B+ie^{-i\varphi _f}(\left|
g_L\right\rangle _A\left| g_L\right\rangle _B+\left| g_R\right\rangle
_A\left| g_R\right\rangle _B)].
\end{equation}
The scheme is deterministic and viable for a rather wide range of
system parameters. It should be noted that the process for the
generation of the qutrit-qutrit entangled state in the present
scheme is at variance with previous strategies adopting dispersive
interactions [27], because here the occupation of both the atomic
excited states {\em and} of the photonic states in the cavities and
fibre are negligible. Moreover, the external control in the present
scheme is less demanding, as the preparatory step to put the atom
in a specific superposition of two ground states [25,27]
or the local manipulation of one of the atoms (by yet another classical field)
during the temporal evolution of the whole system [25],
are not required here.

In the above analysis, exact knowledge of the system parameters is assumed.
But, in general, there could be various errors in the parameters due to the imperfect
characterisation of the system. The potential errors include:
\begin{itemize}
\item{the
mismatch of the coupling rate $g_{x,k}$ and $\Omega _x$ $(x=A,B$; $k=L,R$.$)$
for the atoms with the local cavity and classical fields, as $g_{x,k}$ and $%
\Omega _x$ are dependent on the atomic position and might fluctuate;}
\item{the mismatch of the phase $\phi _x$ due to noise in the phases of the classical
fields;}
\item{the mismatch of the detuning $\Delta _{x,yz}$ ($x=A,B;$ $%
y=e_0,e_L,e_R;$ and $z=g_0,g_L,g_R$) between atoms and fields due to
possibly imprecise control;}
\item{the mismatch of the coupling rate $\nu _k$
for the cavity and fibre modes, as $\nu _k$ is decided by the manufacture
technology and might be imprecise;}
\item{timing errors, due to the finite switching rates of the interactions and
the limited precisions of the interaction times ($t_A$ and $t_B$ will be set for each
atom interacting with the local fields.);}
\item{polarisation errors due
to unstable magnetic fields, which lead to mismatches in two
parameters related to polarisation, such as $g_{x,L(R)}$, $\nu _{L(R)}$ and $\Delta
_{x,L(R)}$ ($\Delta _{x,L(R)}$ here denotes the detuning for $\pi -\sigma
^{\dag}$ or $\pi -\sigma ^{-}$ Raman channel).}
\end{itemize}

\begin{figure}[t!]
\centerline{
\includegraphics[scale=.5]{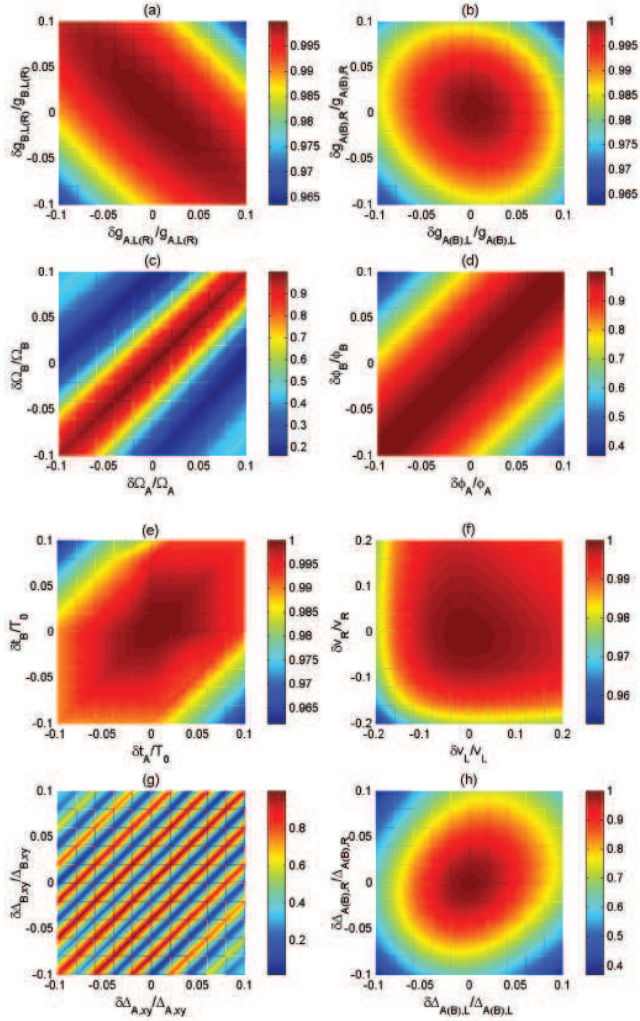}}
\caption{ The fidelity of the qutrit-qutrit entangled state versus
kinds of errors (all the parameters plotted are dimensionless). (a)
$F
$ vs $\frac{\delta g_{A,L(R)}}{g_{A,L(R)}}$ and $\frac{\delta g_{B,L(R)}}{%
g_{B,L(R)}}$; (b) $F$ vs $\frac{\delta g_{A(B),L}}{g_{A(B),L}}$ and $\frac{%
\delta g_{A(B),R}}{g_{A(B),R}}$; (c) $F$ vs $\frac{\delta \Omega
_A}{\Omega _A}$ and $\frac{\delta \Omega _B}{\Omega _B}$; (d) $F$ vs
$\frac{\delta \phi _A}{\phi _A}$ and $\frac{\delta \phi _B}{\phi
_B}$; (e) $F$ vs $\frac{\delta t_A}{T_0}$ and $\frac{\delta
t_B}{T_0}$; (f) $F$ vs $\frac{\delta v_L}{v_L}$
and $\frac{\delta v_R}{v_R}$; (g) $F$ vs $\frac{\delta \Delta _{A,xy}}{%
\Delta _{A,xy}}$ and $\frac{\delta \Delta _{B,xy}}{\Delta _{B,xy}}$;
and (h) $F$ vs $\frac{\delta \Delta _{A(B),L}}{\Delta _{A(B),L}}$
and $\frac{\delta \Delta _{A(B),R}}{\Delta _{A(B),R}}$.}
\end{figure}

In order to check out how the mentioned errors influence the generation of
the entangled state, we define the following fidelity as a measure of the
reliability of the qutrit-qutrit maximally entangled state:
\begin{equation}
F=\left\langle \psi _{AB}^{3D}\right| Tr_{c_1,f,c_2}[\rho (t)]\left| \psi
_{AB}^{3D}\right\rangle ,
\end{equation}
where $\rho (t)$ is the state of the entire system at arbitrary time
(governed by Eq. $(9)$, where neither decoherence nor errors are
accounted for),
and $%
Tr_{c_1,f,c_2}$ denotes the partial trace over the field degrees of freedom.

We first assume ``perfect'' interactions, considering the case $%
g_{x,k}\equiv g$, $\Omega _x=\Omega \equiv g$, $\Delta _{A,e_0g_0}=\Delta
_{B,e_kg_k}\equiv 20g$, $\Delta _{A,e_0g_k}=\Delta _{B,e_kg_0}\equiv 21g$, $%
\phi _A\equiv \phi _B$, and $\nu _k=v\equiv \sqrt{2}g$) as a reference.
Under such conditions maximal qutrit-qutrit
entanglement is obtained at the reference time $t=T_0$ (in the notation of the previous section,
only the case $m=0$ is considered, {\em i.e.,} $\chi T_0\equiv 0.6755$.). We then set
the errors involved in the parameters $g_{x,k}$, $\Omega _x$, $\Delta
_{x,yz} $, $\nu _k$, $\phi _k$ and $t_x$ to be $\delta g_{x,k}$, $\delta
\Omega _x$, $\delta \Delta _{x,yz}$, $\delta v_k$, $\delta \phi _k$ and $%
\delta t_x$, respectively. In Fig.~$2$, the fidelity is plotted
versus the different kinds of errors.
Notice that these fidelity plots display a number of symmetries.
The mirror symmetries about the line bisecting the axes trivially reflect the
choices of the error parameters and the symmetry of the system
under exchange of the two atoms and cavities.
Some of the symmetries are instead more interesting:
for instance, errors in the detunings induce additional phases and have clearly oscillatory
effects (g), errors in the cavity-fibre coupling strengths induce
different Stark shifts (and have hence different effects) depending on their signs (f),
while errors in the atom-light coupling strengths in the two different polarisations
have, more intriguingly, approximately `rotationally symmetric' effects (b).
Let us now quantitatively discuss the influence of the various kinds of errors on the
fidelity with the entangled `reference state'.

It can be seen from Fig.~$2$ $(a)$, $(b)$, $(e)$ and $%
(f)$ that the fidelity $F$ is very robust against errors in the
parameters $g_{x,k}$, $\nu _k$ and $t_x$. A deviation $\left| \delta
g_{x,k}\right| \simeq 10\%$$g_{x,k}$, $\left| \delta \nu _x\right|
\simeq 10\%$$\nu _x$, or $\left| \delta t_x\right| \simeq 10\%$$T_0$
will cause only a reduction smaller than $10^{-2}$ in the fidelity.

From Fig.~$2$ $(c)$, it is apparent that the fidelity is, on the
other hand, very sensitive to imperfections in $\Omega _x$, mainly
dependent on the Stark shifts induced by the classical fields [see
Eq.~$(3)$]. However, the influences of imperfect $\Omega _x$ through
such Stark shifts can be eliminated as one can apply a second
classical field to produce offsetting ac-Stark shifts on both atoms
[27]. If this is done, then the effect of the errors $\left| \delta
\Omega _x\right| $ on $\Omega _x$ will be analogous to the effect of
the deviation $\left| \delta g_{x,k}\right| $ in $g_{x,k}$, which
have already been shown to be very slight. Thus, this simple
countermeasure would make the entanglement fidelity robust also
against possible errors in $\Omega _x$.

When deriving the effective Hamiltonian $(3)$, we set the condition
$\phi _A\equiv \phi _B$ to eliminate the phases of the classical
fields in Eq.~$(3)$ (see Appendix A). However, such phases are
affected by noise and fluctuations, and might be slightly different
in practical instances. Nevertheless, the fidelity is only
marginally degraded by the possible errors in the parameter $\phi
_x$: a deviation $\left| \delta \phi _k\right| =3\%$$\phi _k$ will
cause only $2\%$ in the reduction of the fidelity. In practice,
Using only one classical field to illuminate both atoms that are
distributed in the two cavities would help in keeping the phase
fluctuations under control [31].

Let us remind the reader that in our scheme the cavity and the
classical fields are detuned from the corresponding atomic
transitions by specific values. It can be seen from Fig.~$2$ $(g)$
that the fidelity is highly dependent on the parameter $\Delta
_{x,yz}$: a small deviation in $\Delta _{x,yz}$ leads to large
oscillation in the fidelity. This is mainly due to the
detuning-dependent Stark shifts induced by the cavity and classical
fields, besides the possible occurrence of a phase ($\Delta =\Delta
_{A,e_0g_k}-\Delta _{A,e_0g_0}-\Delta _{B,e_kg_0}+\Delta
_{B,e_kg_k}) $ in the exponential factor $e^{i\Delta t}$ in
Eq.~$(3)$. Though this requirement is strict, it is not a major
problem with the currently developed laser technology in cavity QED
experiments: the necessary stabilisation of the fields' frequencies
can be achieved by means of acousto--optic modulators [32]. In
Fig.~$2$ $(h)$, we
check the stability of the fidelity versus the detunings for the two polarisation (%
$\pi -\sigma ^{\dag}$ or $\pi -\sigma ^{-})$ Raman channels.
Ideally, the conditions $\Delta _{A,L}$=$\Delta _{A,R}$ and $\Delta
_{B,L}$=$\Delta _{B,R}$ are required. But, in real experiments, this
requirement may not be perfectly satisfied, due to the fact that the
magnetic field can break the degeneracy between the atomic ground
states. Our investigation, reported in Fig.~$2$ $(h)$ shows that the
fidelity is only slightly affected by errors in the detunings for
both polarisation channels: the fidelity with the entangled state of
reference will still be larger than $0.96$ even when a deviation
$\left| \delta \Delta _{A(B),k}\right| \simeq 3\%$$\Delta _{A(B),k}$
occurs.

\begin{figure}[t!]
\centerline{
\includegraphics[scale=.5]{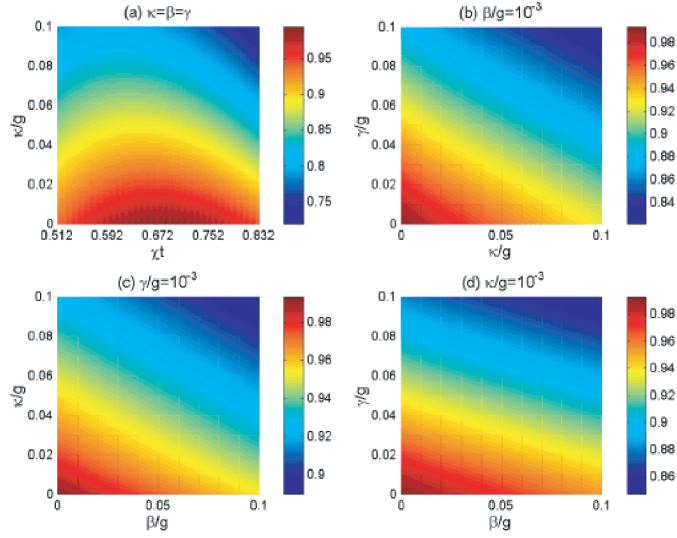}}
\caption{ The fidelity of the qutrit-qutrit entangled state
versus the dimensionless parameters $\chi t$, $\kappa /g$, $\gamma /g$ or $%
\beta /g$. (a) $F$ vs $\chi t$ and $\kappa /g$ ($\kappa =\beta
=\gamma $);
(b) $F$ vs $\kappa /g$ and $\gamma /g$ ($\beta =10^{-3}g$); (c) $F$ vs $%
\gamma /g$ and $\beta /g$ ($\kappa =10^{-3}g$); and (d) $F$ vs
$\kappa /g$ and $\beta /g$ ($\gamma /g=10^{-3}g$).}
\end{figure}

\section{Influence of spontaneous emission and photon leakage}\label{noise}

In all the above arguments, we have assumed the entire system is ideally
isolated from the environment, and have not considered any dissipation.
In this section, we take into account the dissipation due to
atomic spontaneous emission and photon leakage from the cavities and fibre.
The master equation for the density matrix of the entire system
can be expressed as
\begin{eqnarray}
\fl \hspace*{1cm} \dot\rho &=&-i[H_{full},\rho ]+\frac \kappa 2%
\sum_{k=L,R}[\sum_{x=A,B}(2a_{x,k}\rho a_{x,k}^{\dag}-a_{x,k}^{\dag}a_{x,k}\rho
-\rho a_{x,k}^{\dag}a_{x,k})  \nonumber \\
\fl&& +\frac \beta 2(2b_k\rho b_k^{\dag}-b_k^{\dag}b_k\rho -\rho
b_k^{\dag}b_k)]  \nonumber \\
\fl && +\frac \gamma 2\sum_{x=A,B}\sum_{\sigma =L,R,\pi
}(2A_{x,\sigma }\rho A_{x,\sigma }^{\dag}-A_{x,\sigma }^{\dag}A_{x,\sigma }\rho
-\rho A_{x,\sigma }^{\dag}A_{x,\sigma }),
\end{eqnarray}
where $A_{x,\sigma }=\sum_{y,z}\left| y\right\rangle _x\left\langle y;1\sigma \mid z\right\rangle _x\left\langle z\right| $ ($y=g_L$, $g_0$, $g_R$%
; $z=e_L$, $e_0$, $e_R$.) is the atomic lowering operator, with $%
_x\left\langle y;1\sigma \mid z\right\rangle _x$ being the
Clebsch-Gordan coefficient $($i.e., $C_{m,m^{\prime }})$ for the
dipole transition $\left| e\right\rangle \leftrightarrow \left|
g\right\rangle $ with polarisation $\sigma =L$, $R$, $\pi $; $\gamma
$, $\beta $ and $\kappa $ stand, respectively, for the spontaneous
emission rate and for the fibre and cavity decay rates (assumed for
simplicity to be equal for the two cavities and for the two
polarised modes). The contribution of the thermal photons have been
neglected, as is possible at optical frequencies.

The master equation $(9)$ has been numerically solved in the
subspace $\Gamma \in \{\Gamma _{full}$, $\left| g_L\right\rangle
_A\left| g_0\right\rangle _B\left| 00\right\rangle _{c_1}\left|
00\right\rangle _{fib}\left| 00\right\rangle _{c_2}$, $\left|
g_R\right\rangle _A\left| g_0\right\rangle _B\left| 00\right\rangle
_{c_1}\left| 00\right\rangle _{fib}\left| 00\right\rangle _{c_2}\}$.
In Fig.~$3$ $(a)$, the fidelity of the maximal qutrit-qutrit
entangled state is plotted versus the dimensionless parameters $\chi t$ and $%
\kappa /g$ ($\kappa =\beta =\gamma $ is set).
In Fig.~$3$ $(b)$, $(c)$ and $(d)$ the fidelity is plotted versus each pair of the three
dimensionless parameters
$\kappa /g$, $\beta /g$ and $\gamma /g$ (the remaining one is set to be $%
10^{-3}g$). In the calculations, we still set $\Omega _A=\Omega
_B=g_{A,k}=g_{B,k}\equiv g$, $\Delta _{A,e_0g_0}=\Delta
_{B,e_kg_k}\equiv 20g $, $\Delta _{A,e_0g_k}=\Delta
_{B,e_kg_0}\equiv 21g$, and $\nu _k\equiv \sqrt{2}g$.

From Fig.~$3$ $(a)$, we note that the fidelity is almost
unaffected by the three decay rates $\kappa $, $\beta $ and $\gamma $ when $%
\kappa =\beta =\gamma =10^{-3}g$. Even when $\kappa =\beta =\gamma =10^{-2}g$%
, the fidelity is close to $0.97$, which is much larger than the one
($<0.87$) obtained in Ref.~[27]. This improvement is of course due
to the suppression of the excited states' population of the fields,
as well as of the atoms. From Fig.~$3$ $(b)$, $(c)$ and $(d)$, it
can be seen that a decay rate of $10^{-2}g$, for either $\kappa$,
$\beta$ or $\gamma$ alone (with the other two parameters set to
zero) leads to a fidelity larger than $0.98$. Note that
the previous scheme through the adiabatic passage [25], a decay rate $%
\kappa \equiv 10^{-2}g$ alone degraded the fidelity down to $F=0.95$.

\section{Conclusion}\label{conclu}

In summary, we have proposed a scheme of atomic levels (with an explicit
possible realisation in Zeeman sublevels of alkali atoms), where
qutrit quantum information can be stored in three ground states and, most importantly,
manipulated globally between distant nodes through the virtual excitation of
excited atomic levels and mediating bosonic fields like, typically, light.

Our scheme is different from any previously proposed ones in that
this choice of atomic levels allows for the whole coherent evolution of the global system
-- involving the two atoms and the linking bosonic modes --
to be driven by the virtual excitation of {\em both} the atomic excited levels
{\em and} the intervening fields.

This feature renders our scheme remarkably more robust than any other
previously proposed in the face of decoherence, whose main sources in these settings are photon loss and spontaneous emission from excited levels.
Also, our scheme -- being based, essentially, on the proper choice of atomic levels and operating regimes -- requires very modest control and proves to be rather resilient against
experimental imperfections as well.
All these qualitative remarks have been substantiated in this work
by a very thorough quantitative analysis of such unwanted effects.

Clearly, a price has to be paid for improved robustness: the use of
exclusively virtual excitations makes these coherent manipulations
very slow if compared to schemes adopting resonant couplings [33].
Ultimately, the choice between faster, resonant schemes and more
robust, virtual ones should depend on the use one intends to make of
them. Of course, speed would be paramount in applications directly
related to quantum computation. However, as pointed out in the
introduction, qutrit systems are mainly interesting for quantum
communication purposes, where the most delicate task to accomplish
is precisely the robust distribution of entanglement between distant
nodes of a network, and speed is not as crucial: the present, fully
virtual scheme would respond precisely to this need. In this
perspective, our study shows that atomic systems hold considerable
promise for the encoding and coherent, distributed manipulation of
multidimensional quantum alphabet for quantum communication
purposes.


\section*{Appendix A: Derivation of the effective Hamiltonian}
Under the condition of large detuning, {\em i.e.} for $\Delta
_{x,yz}\gg $ $g_{x,k}$, $\Omega _x$, and as long as the atoms are
initialised in the ground states, the probability that the atomic
excited states are populated is virtual. Thus, the atomic excited
states are negligible during the time evolution of the entire
system. In this case, we can adopt the time-averaging method [34] to
obtain an effective Hamiltonian as follows [35]:.

\begin{eqnarray}
H_e^{acl} &=&-iH_I^{acl}(t)\int H_I^{acl}(t^{\prime })dt^{\prime }
\nonumber
\\
\ &=&\sum_{k=L,R}[\mu _{\Omega _A}\left| g_0\right\rangle
_A\left\langle g_0\right| +\mu _{\Omega _B}\left| g_k\right\rangle
_B\left\langle g_k\right|
\nonumber \\
&&\ +\mu _{a_{A,k}}a_{A,k}^{\dag}a_{A,k}\left| g_k\right\rangle
_A\left\langle g_k\right| +\mu _{a_{B,k}}a_{B,k}^{\dag}a_{B,k}\left|
g_0\right\rangle
_B\left\langle g_0\right|  \nonumber \\
&&\ +\lambda _{a_{A,k}\Omega _A}(a_{A,k}e^{i\Delta _{A,k}t}e^{-i\phi
_A}\left| g_0\right\rangle _A\left\langle g_k\right| +H.c)  \nonumber \\
&&\ +\lambda _{a_{B,k}\Omega _B}(a_{B,k}e^{i\Delta _{B,k}t}e^{-i\phi
_B}\left| g_k\right\rangle _B\left\langle g_0\right| +H.c)],
\end{eqnarray}
where $\mu _{\Omega _A}=\frac{\Omega _A^2}{\Delta _{A,e_0g_0}}$,
$\mu
_{\Omega _B}=\frac{\Omega _B^2}{\Delta _{B,e_kg_k}}$, $\mu _{a_{A,k}}=\frac{%
g_{A,k}^2}{\Delta _{A,e_0g_k}}$, $\mu
_{a_{B,k}}=\frac{g_{B,k}^2}{\Delta
_{B,e_kg_0}}$, $\lambda _{a_{A,k}\Omega _A}=\frac{g_{A,k}\Omega _A}2(\frac 1{%
\Delta _{A,e_0g_0}}+\frac 1{\Delta _{A,e_0g_k}})$, $\lambda
_{a_{B,k}\Omega _{B,k}}=\frac{g_{B,k}\Omega _B}2(\frac 1{\Delta
_{B,e_kg_0}}+\frac 1{\Delta
_{B,e_kg_k}})$, $\Delta _{A,k}=\Delta _{A,e_0g_k}-\Delta _{A,e_0g_0}$, $%
\Delta _{B,k}=\Delta _{B,e_kg_0}-\Delta _{B,e_kg_k}$. For
$H_e^{acl}$ in Eq.~(10), the first (second) and third (fourth) terms
describe the Stark shifts for the states $\left| g_0\right\rangle $
($\left| g_k\right\rangle $) and $\left| g_k\right\rangle $ ($\left|
g_0\right\rangle $) of the atom in cavity A (B), induced by the
classical and cavity fields, respectively; the fifth (sixth) term
describes the Raman coupling between the states $\left|
g_0\right\rangle $ and $\left| g_k\right\rangle $ for the atom in
cavity A (B), induced collectively by the classical and cavity
fields.

Hence, the effective Hamiltonian of the entire system is given by $%
H_e=H_I^{cf}+H_e^{acl}$.
Let us now introduce three normal modes $c_k$ and $%
c_{\pm k}$ by applying the canonical transformation $c_k=\frac 1{\sqrt{2}}%
(a_{A,k}-e^{-i\varphi _f}a_{B,k})$ and $c_{\pm k}=\frac 12%
(a_{A,k}+e^{-i\varphi _f}a_{B,k}\pm \sqrt{2}b_k)$ [22]. Then, we
switch to a rotating frame by the unitary transformation
$R=e^{-iH_I^{cf}t}$ [29], {\em i.e.}, $H_e^{\prime
}=R^{\dag}H_eR-iR^{\dag}\frac{dR}{dt}$, and obtain

\begin{eqnarray}
\fl H_e^{\prime } &=&\sum_{k=L,R}\{\mu _{\Omega _A}\left|
g_0\right\rangle _A\left\langle g_0\right| +\mu _{\Omega _B}\left|
g_k\right\rangle
_B\left\langle g_k\right|  \nonumber \\
\fl &&\ +\frac{\mu _{a_{A,k}}}4(c_{+k}^{\dag}c_{+k}+c_{-k}^{\dag}c_{-k}+2c_k^{\dag}c_k)%
\left| g_k\right\rangle _A\left\langle g_k\right|  \nonumber \\
\fl &&\ +\frac{\mu _{a_{B,k}}}4(c_{+k}^{\dag}c_{+k}+c_{-k}^{\dag}c_{-k}+2c_k^{\dag}c_k)%
\left| g_0\right\rangle _B\left\langle g_0\right|  \nonumber \\
\fl &&\ +\frac{\mu _{a_{A,k}}}4(c_{+k}^{\dag}c_{-k}e^{i2\sqrt{2}\nu _kt}+\sqrt{2}%
c_{+k}^{\dag}c_ke^{i\sqrt{2}\nu
_kt}+\sqrt{2}c_{-k}^{\dag}c_ke^{-i\sqrt{2}\nu
_kt}+H.c)\left| g_k\right\rangle _A\left\langle g_k\right|  \nonumber \\
\fl &&\ +\frac{\mu _{a_{B,k}}}4(c_{+k}^{\dag}c_{-k}e^{i2\sqrt{2}\nu _kt}-\sqrt{2}%
c_{+k}^{\dag}c_ke^{i\sqrt{2}\nu
_kt}-\sqrt{2}c_{-k}^{\dag}c_ke^{-i\sqrt{2}\nu
_kt}+H.c)\left| g_0\right\rangle _B\left\langle g_0\right|  \nonumber \\
\fl &&\ +\frac{\lambda _{a_{A,k}\Omega _A}}2[(c_{+k}e^{-i\sqrt{2}\nu
_kt}+c_{-k}e^{i\sqrt{2}\nu _kt}+\sqrt{2}c_k)e^{i\Delta
_{A,k}t}e^{-i\phi
_A}\left| g_0\right\rangle _A\left\langle g_k\right| +H.c]  \nonumber \\
\fl &&\ +\frac{\lambda _{a_{B,k}\Omega _B}}2[(c_{+k}e^{-i\sqrt{2}\nu
_kt}+c_{-k}e^{i\sqrt{2}\nu _kt}-\sqrt{2}c_k)e^{i\Delta
_{B,k}t}e^{-i(\phi _B-\varphi _f)}\left| g_k\right\rangle
_B\left\langle g_0\right| +H.c]\}. \nonumber
\end{eqnarray}

For simplicity, we now set $\mu _1=\mu _{\Omega _A}=\mu _{\Omega
_B}$, $\mu _2=\mu _{a_{A,k}}=\mu _{a_{B,k}}$, $\lambda =\lambda
_{a_{A,k}\Omega _A}=\lambda _{a_{B,k}\Omega _B}$, $\Delta =\Delta
_{A,k}=\Delta _{B,k}$, $\phi _A=\phi _B$ and $\nu =\nu _k$. Under
the
condition $\sqrt{2}\nu $, $\left| \Delta -\sqrt{2}\nu \right| $, $\Delta +%
\sqrt{2}\nu $, and $\Delta \gg \frac{\mu _2}4$, $\frac \lambda 2$,
the energy exchange between the bosonic modes and the atoms as well
as between the bosonic modes themselves is virtual. The virtual
excitation of the bosonic modes leads to the Stark shifts and
coupling between the atoms. Then $H_e^{\prime }$ reduces to [29]
\begin{eqnarray}
\fl H_e^{\prime \prime } &=&\sum_{k=L,R}\{\mu _1(\left| g_0\right\rangle
_A\left\langle g_0\right| +\left| g_k\right\rangle _B\left\langle
g_k\right|
)  \nonumber \\
\fl &&\ +\frac{\mu
_2}4(c_{+k}^{\dag}c_{+k}+c_{-k}^{\dag}c_{-k}+2c_k^{\dag}c_k)(\left|
g_k\right\rangle _A\left\langle g_k\right| +\left| g_0\right\rangle
_B\left\langle g_0\right| )  \nonumber \\
\fl &&\ +\frac{\mu _2^2}{32\sqrt{2}\nu }(c_{+k}^{\dag}c_{+k}-c_{-k}^{\dag}c_{-k})(%
\left| g_k\right\rangle _A\left\langle g_k\right| +\left|
g_0\right\rangle
_B\left\langle g_0\right| )^2  \nonumber \\
\fl &&\ +\frac{\mu _2^2}{8\sqrt{2}\nu }(c_{+k}^{\dag}c_{+k}-c_{-k}^{\dag}c_{-k})(%
\left| g_k\right\rangle _A\left\langle g_k\right| -\left|
g_0\right\rangle
_B\left\langle g_0\right| )^2  \nonumber \\
\fl &&\ +\frac{\lambda ^2}4[(\frac 1{\Delta -\sqrt{2}\nu
}c_{+k}c_{+k}^{\dag}+\frac 1{\Delta +\sqrt{2}\nu
}c_{-k}c_{-k}^{\dag}+\frac 2\Delta c_kc_k^{\dag})(\left| g_0\right\rangle
_A\left\langle g_0\right| +\left| g_k\right\rangle
_B\left\langle g_k\right| )  \nonumber \\
\fl &&\ -(\frac 1{\Delta -\sqrt{2}\nu }c_{+k}^{\dag}c_{+k}+\frac 1{\Delta +\sqrt{2}%
\nu }c_{-k}^{\dag}c_{-k}+\frac 2\Delta c_k^{\dag}c_k)(\left|
g_k\right\rangle _A\left\langle g_k\right| +\left| g_0\right\rangle
_B\left\langle g_0\right|
)]  \nonumber \\
\fl &&-\frac{\lambda ^2}4(-\frac 1{\Delta -\sqrt{2}\nu }-\frac 1{\Delta +\sqrt{2}%
\nu }+\frac 2\Delta )(e^{-i\varphi _f}S_{A,k}^{\dag}S_{B,k}^{-}+H.c)\},
\end{eqnarray}
with $S_{A,k}^{\dag}=\left| g_0\right\rangle _A\left\langle g_k\right| $ and $%
S_{B,k}^{-}=\left| g_0\right\rangle _B\left\langle g_k\right| $. The
quantum-number operators $c_{+k}^{\dag}c_{+k}$, $c_{-k}^{\dag}c_{-k}$,
$c_k^{\dag}c_k$ for the bosonic modes are conserved quantities during
the interaction as all of them commute with the Hamiltonian
$H_e^{\prime \prime }$. Suppose all the
modes of the cavities and fibre are initially in the vacuum state, i.e., $%
\left| 00\right\rangle _{c_1}\left| 00\right\rangle _{fib}\left|
00\right\rangle _{c_2}$. Hence, all the bosonic modes $c_{+k}$, $c_{-k}$ and $%
c_k$ will remain in the vacuum state during the evolution.

Finally, the global effective Hamiltonian $H_e^{\prime \prime }$
reads
\begin{equation}
\fl H_e^{\prime \prime }=\sum_{k=L,R}\eta (\left| g_0\right\rangle
_A\left\langle g_0\right| +\left| g_k\right\rangle _B\left\langle
g_k\right| )-\chi (e^{-i\varphi _f}S_{A,k}^{\dag}S_{B,k}^{-}+H.c),
\end{equation}
where
\begin{equation}
\eta =\mu _1+\eta' ,
\end{equation}
\begin{equation}
\eta' =\frac{\lambda ^2}4[(\frac 1{\Delta -\sqrt{2}\nu }+\frac 1{\Delta +%
\sqrt{2}\nu }+\frac 2\Delta ),
\end{equation}
and
\begin{equation}
\chi =\frac{\lambda ^2}4(-\frac 1{\Delta -\sqrt{2}\nu }-\frac 1{\Delta +%
\sqrt{2}\nu }+\frac 2\Delta ).
\end{equation}
It should be noted that we have neglected the term $\eta' \left|
g_0\right\rangle _A\left\langle g_0\right| $ to maintain the
symmetry in the effective Hamiltonian $(12)$. In practice, this term
can be compensated by an
additional ac-Stark shift for the state $\left| g_0\right\rangle $ of atom $%
A $ [27].

\section*{Appendix B: Validity of the effective dynamics}

We now turn back to the full Hamiltonian of the system and check how
accurate is the description of the system through the effective
Hamiltonian $(3)$.
We take the energy level $\left| F=2\right\rangle
$ of $5^2S_{1/2}$ to be the zero energy reference point, and write down
the full Hamiltonian for the entire system as follows:
\begin{eqnarray}
\fl H_{full} &=&\sum_{k=L,R}(\omega _{f,k}b_k^{\dag}b_k+\omega
_{a_{A,k}}a_{A,k}^{\dag}a_{A,k}+\omega _{a_{B,k}}a_{B,k}^{\dag}a_{B,k}
\nonumber
\\
\fl &&\omega _{A,g_k}\left| g_k\right\rangle _A\left\langle g_k\right|
+\omega _{B,e_k}\left| e_k\right\rangle _B\left\langle e_k\right|
)+\omega _{A,e_0}\left| e_0\right\rangle _A\left\langle e_0\right|
+\omega
_{B,g_0}\left| g_0\right\rangle _B\left\langle g_0\right|  \nonumber \\
\fl &&+\sum_{k=L,R}[g_{A,k}a_{A,k}\left| e_0\right\rangle _A\left\langle
g_k\right| +\Omega _Ae^{-i(\omega _{\Omega _A}t-\phi _A)}\left|
e_0\right\rangle _A\left\langle g_0\right|  \nonumber \\
\fl &&+g_{B,k}a_{B,k}\left| e_k\right\rangle _B\left\langle g_0\right|
+\Omega _Be^{-i(\omega _{\Omega _B}t-\phi _B)}\left|
e_k\right\rangle _B\left\langle
g_k\right| +H.c]  \nonumber \\
\fl &&+\sum_{k=L,R}\nu _k[b_k(a_{A,k}^{\dag}+e^{i\varphi
_f}a_{B,k}^{\dag})+H.c],
\end{eqnarray}
where $\omega _{f,k}$ and $\omega _{a_{A,k}}$ ($\omega _{a_{B,k}}$)
are the energy levels for the polarised photons in the fibre and
cavity A (B),
respectively, $\omega _{\Omega _x}$ ($x=A,B$) denotes the energy for the $%
\pi $-polarised classical field $\Omega _x$, and $\omega _{x,m}$ ($m=g_0$, $%
g_k$, $e_0$, $e_k$) is the energy level for the atomic state $\left|
m\right\rangle _x$.

Taking the initial state $\left| \psi (0)\right\rangle =\left|
g_0\right\rangle _A\left| g_0\right\rangle _B\left| 00\right\rangle
_{c_1}\left| 00\right\rangle _{fib}\left| 00\right\rangle _{c_2}$
and considering all possible states of the system in evolution, we
express the state of the system at time $t$ as $\left| \psi
_{full}(t)\right\rangle =\sum_{i}c_i(t)\left| \phi _i\right\rangle $
($c_i(t)$ being time-dependent amplitudes)
within the subspace $\Gamma_{full}$ spanned by the vectors
$\{\left| \phi
_1\right\rangle , \ldots, \left| \phi _i\right\rangle , \ldots,
\left| \phi _{12}\right\rangle \}$:
\begin{eqnarray}
\fl \Gamma_{full}&\equiv&
\{(\left| g_0\right\rangle
_A\left| g_0\right\rangle _B, \left| e_0\right\rangle _A\left|
g_0\right\rangle _B, \left| g_L\right\rangle _A\left|
g_L\right\rangle _B, \left| g_R\right\rangle _A\left| g_R\right\rangle _B
, \left| g_L\right\rangle _A\left| e_L\right\rangle _B, \left|
g_R\right\rangle _A\left| e_R\right\rangle _B) \nonumber \\
\fl &&\otimes \left|
00\right\rangle _{c_1}\left| 00\right\rangle _{fib}\left|
00\right\rangle _{c_2}, \left| g_L\right\rangle _A\left|
g_0\right\rangle _B \nonumber \\
\fl &&\otimes (\left| 10\right\rangle _{c_1}\left|
00\right\rangle _{fib}\left| 00\right\rangle _{c_2}, \left|
00\right\rangle _{c_1}\left| 10\right\rangle _{fib}\left|
00\right\rangle _{c_2}, \left| 00\right\rangle _{c_1}\left|
00\right\rangle _{fib}\left| 10\right\rangle _{c_2}), \left|
g_R\right\rangle _A\left| g_0\right\rangle _B \nonumber \\
\fl && \otimes (\left|
01\right\rangle _{c_1}\left| 00\right\rangle _{fib}\left|
00\right\rangle _{c_2}, \left| 00\right\rangle _{c_1}\left|
01\right\rangle _{fib}\left| 00\right\rangle _{c_2}, \left|
00\right\rangle _{c_1}\left| 00\right\rangle _{fib}\left|
01\right\rangle _{c_2})\}\, . \nonumber
\end{eqnarray}
The occupation probability for each state vector $\left|
\phi _i\right\rangle $ during the
evolution is $P_i(t)=\left| c_i(t)\right| ^2$, and satisfies $%
\sum_{i=1}^{12}P_i(t)=1$. Thus the occupation probability of the
atomic
excited states and the photonic states are $P_e(t)=\sum_{i=2,5,6}P_i$ and $%
P_p(t)=\sum_{i=7}^{12}P_i$, respectively. The validity of the
effective Hamiltonian implies that both the occupation probability
$P_e(t)$ and $P_p(t)$ should be small enough thus they can be
negligible during the time evolution of the entire system. We focus
here on quantum state transfer, {\em i.e.}, on the variation of the
occupation probability $P_1(t)$ and $P_{tra}(t)$ $\equiv
\sum_{i=3,4}P_i$, for the numerical verification of the effective dynamics,
which is portrayed in Fig.~$4$.
\begin{figure}[t!]
\centerline{
\includegraphics[scale=.5]{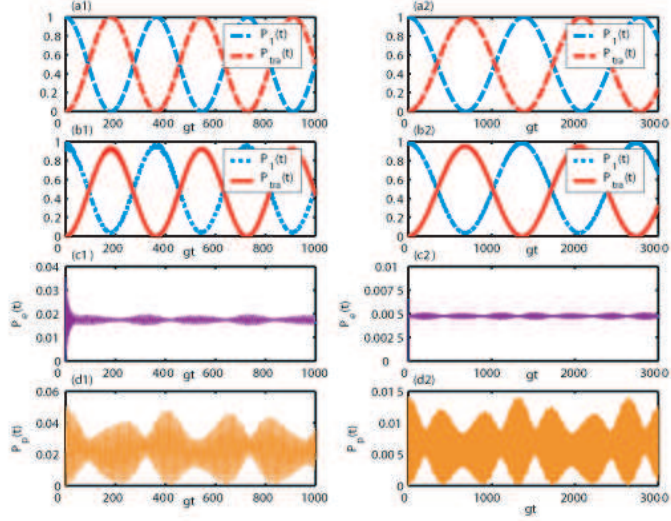}}
\caption{The occupation probability $P_1(t)$, $P_{tra}(t)$%
, $P_e(t)$ and $P_p(t)$ versus the dimensionless parameter $gt$,
respectively, $\Omega _A=\Omega _B=g_{A,k}=g_{B,k}\equiv g$, $\Delta
_{A,e_0g_0}=\Delta _{B,e_kg_k}=\Delta _1$, $\Delta
_{A,e_0g_k}=\Delta _{B,e_kg_0}=\Delta _2$ and $\nu \equiv
\sqrt{2}g$. $(a1)\sim (d1)$ $\Delta
_1\equiv 10g$, $\Delta _2\equiv 11g$; $(a2)\sim (d2)$ $\Delta _1=20g$, $%
\Delta _2=21g$.}
\end{figure}
Fig.~$4$ $(a1)$ and $(a2)$ are
obtained through the solution of the Schr\"odinger equation
$i\frac{d\left| \psi _{eff}(t)\right\rangle }{dt}=H_e^{\prime \prime
}\left| \psi _{eff}(t)\right\rangle $ in the subspace $\Gamma _e\in
\{\left| \phi _1\right\rangle $, $\left| \phi _3\right\rangle $,
$\left| \phi _4\right\rangle \}$. The two figures display perfect
Rabi oscillations, which indicates ideal state transfer between the
states $\left| \phi _1\right\rangle $ and $1/\sqrt{2}(\left| \phi
_3\right\rangle +\left| \phi
_4\right\rangle )$.

Fig.~$4$ $(b1)$, $(c1)$, $(d1)$, $(b2)$, $(c2)$ and $%
(d2)$ are obtained through the solution of the Schr\"odinger equation $i%
\frac{d\left| \psi _{full}(t)\right\rangle }{dt}=H_{full}\left| \psi
_{full}(t)\right\rangle $ in the subspace $\Gamma _{full}$. It can
be seen from Fig.~$4$ $(b1)$ and $(b2)$ that state transfer via the
effective Hamiltonian $(3)$ is almost perfect (for Fig.~$4$ $(b2)$,
this is especially apparent.), indicating that numerical results
obtained from the effective and full Hamiltonians would be
equivalent. Fig.~$4$ $(c1)$ and $(c2)$ plot the variation of the
occupation probability of the atomic excited states, while Fig.~$4$
$(d1)$ and $(d2)$ plot the variation of the occupation probability
of the photonic states in the cavities and fibre. It is apparent
that both $P_e$ and $P_p$ are very small during the evolution of the
entire system. The analysis above verifies the validity of the
effective Hamiltonian $(3)$.

Let us now review the physical conditions given in the description of our system,
to reveal some insight about and relationships
between certain dynamical parameters when the effective Hamiltonian $(3)$ is valid.
In order for $H_e^{\prime \prime }$ to hold, we required
$\Delta _i\gg g $, $\Omega $ as well as $\sqrt{2}\nu $, $\left|
\Delta \pm \sqrt{2}\nu \right| $, $\Delta \gg \frac{g^2}{4\Delta
_i}$, $\frac{g\Omega }{2\Delta _i}$ ($\Delta \ll \Delta _i$,
$i=1,2$). $\Delta _i$ is the dominant factor, because the occupation
probability of the atomic excited states $(P_e)$ and the photonic
states $(P_p)$ are inversely proportional to $\Delta _i^2$,
given all other parameters are pre-set. This is proved to be true in Fig.~$%
4 $ $(c1)$, $(d1)$, $(c2)$, and $(d2)$, as the average occupation
probability $P_e$ and $P_p$ shown in Fig.~$4$ $(c1)$ and $(d1)$
are about four times that in Fig.~$4$ $(c2)$ and $(d2)$.
In other words, the
difference between the effective Hamiltonian $(3)$ and the full Hamiltonian $%
(16)$ decreases with increasing $\Delta _i$. This can also explain
the phenomena for the different deviation from the perfect state
transfer, which are more noticeable in Fig.~$4$ $(b1)$ than in
Fig.~$4$ $(b2)$.

Let us stress once again that the obtained effective Hamiltonian $(3)$ is
indeed valid as the occupations of the atomic excited states and the
photonic states have been showed to be strongly suppressed.

\subsection*{Acknowledgments}

This work is supported by National Natural Science Foundation of
China under Grant No. 10674025 and No. 10974028, the Fujian Natural
Science Foundation of China under Grant No. 2009J06002, Doctoral
Foundation of the Ministry of Education of China under Grant No.
20070386002, funds from State Key Laboratory Breeding Base of
Photocatalysis, Fuzhou University, and funds from Education
Department of Fujian Province of China under Grant No. JB08010. SYY
is supported by a KC Wong Scholarship. AS thanks the Central
Research Fund of the University of London for financial support.

\end{document}